# Plasmonic Crystals with Tunable Band Gaps in the Grating Gate Transistor Structures

G. R. Aizin[1,†], J. Mikalopas[1], and M. Shur[2,3]

[1] *Kingsborough College, The City University of New York, Brooklyn, New York 11235, USA*
[2] *Rensselaer Polytechnic Institute, Troy, New York 12180, USA*
[3] *Electronics of the Future, Inc., Vienna, VA 22181, USA*

We developed a hydrodynamic model of plasmonic crystals formed in the current-driven grating gate transistor structures. The model demonstrates that the quality factor of plasmonic resonances could be increased by using ungated regions with high electron densities connecting multiple plasmonic cavities. The analytical and numerical calculations of the EM radiation absorption by the band plasmons show that the drive current makes all plasma modes optically active by breaking the symmetry of the plasma oscillations. This effect results in splitting plasmon resonant absorption peaks revealing the gaps in the plasmonic band spectrum tunable by current. The analyzed design could achieve resonant behavior at room temperature for plasmonic crystals implemented in various material systems, including graphene, III-V, III-N materials, and p-diamond. We argue that the resulting double-peak spectrum line in the terahertz range also facilitates the absorption at the gap frequency, typically in microwave range. Power pumping at the gap frequency enables excitation of the gap plasmons, promoting frequency conversion from microwave to THz ranges. The flexibility in the length of the ungated region for the investigated structures allows for an effective coupling with THz radiation, with the metal grating acting as a distributive resonant antenna. The applications of the presented results extend to THz communication systems, THz sensing and imaging, frequency conversion systems, and other advanced THz plasmonic devices.

## I. INTRODUCTION.

The lateral plasmonic crystals are formed in two-dimensional (2D) electron channels of the field-effect transistors periodically modulated in space when the electron mean free path is longer than several modulation periods [1]. This condition can be met even at room temperature in advanced semiconductor material systems with long mean free path such as AlGaAs/InGaAs [2], AlGaN/GaN [3], p-diamond [4] or in graphene [5]. Scaling down the dimensions of semiconductor devices below 10 nm stimulated theoretical and experimental research of resonant plasmonic crystals in periodically modulated 2D electron systems [6-24].

Periodic modulation of the 2D electron channel induces gap openings in the plasmon energy spectrum. This effect was first described theoretically in [25] and confirmed experimentally in [26]. Recently it got renewed interest due to its potential applications in the sub-THz and THz technology [6]. Frequencies of 2D plasmons in the THz band are easily tunable in the THz field-effect transistors (TeraFETs) by the gate bias or illumination, and the wide-ranging applications of the plasmonic TeraFETs in the THz electronics as compact tunable detectors and sources of the THz EM radiation are anticipated [27, 28]. The advantages of plasmonic crystal over plasmons in a single plasmonic cavity transistor are better coupling with an external EM wave [7] and signal amplification due to coherence of the plasma oscillations in individual elementary cells [1] resulting in more sensitive detectors and efficient THz sources.

Systematic studies of the plasmonic crystal effects in periodically modulated TeraFETs began more than a decade ago [7-13]. Plasmonic band structure in the finite plasmonic crystal formed in the TeraFET with periodically modulated electron density was demonstrated experimentally [8] and described theoretically including Tamm plasmonic states formed at the edges of the finite plasmonic crystal [9]. Plasmonic crystals in graphene structures with periodically modulated geometry were studied in [10-12].

In the presence of a DC current bias, plasmonic band spectra are modified [13], and plasmonic crystals develop plasmonic boom instability if the electron drift velocity exceeds the plasma velocity [14]. This effect is similar to the sonic boom instability in acoustics [29]. The THz emission due to plasmonic boom effect was observed in the grating gate TeraFETs [15]. Driven by DC current bias, plasmonic crystals may also develop Dyakonov-Shur type of instability due to current induced asymmetry in the plasmonic crystal elementary cell [16] and/or the built-in asymmetry of the elementary cell [17]. Recently, an amplified mode switching (AMS) effect has been observed in the current biased

---

† gaizin@kbcc.cuny.edu

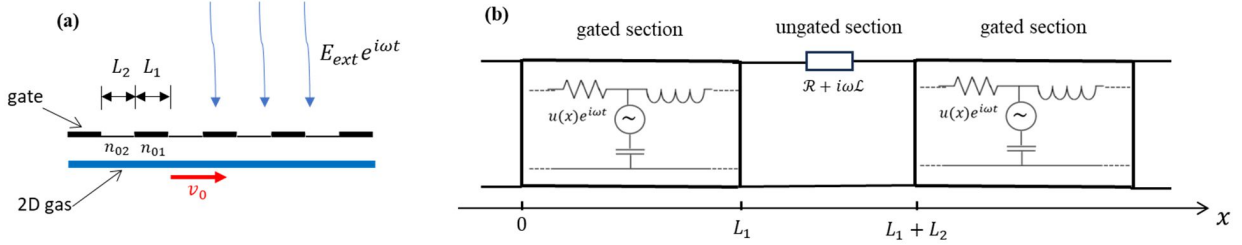

Fig.1 (a) Schematics of the current biased TeraFET structure irradiated by an external EM wave at normal incidence; (b) Equivalent electric circuit representing TeraFET's 2D electron channel.

interdigitated graphene plasmonic crystals [18]. It occurs at certain critical value of the ratio of drift velocity over plasma velocity (the Mach number) when the plasma resonant peak experiencing redshift with increasing current changes to the blue shifting peak. The AMS effect was explained by switching between crossed plasmonic modes with different effective plasmonic damping [19].

Absorption of impinging THz EM radiation in the plasmonic crystal can be used for the detection of THz signals. This effect was recently demonstrated experimentally in the grating-gated AlGaN/GaN quantum well nanostructures [20] where absorption peaks tunable by the gate voltage were observed and explained by the formation of a plasmonic crystal. The detailed theory of absorption of the THz radiation in the plasmonic crystals with modulated electron density controlled by the gate was developed in [21, 22].

In this paper, we consider plasmonic crystals in the grating gated TeraFET in which the unit cell consists of gated and ungated regions with electron density and the plasma frequency in the ungated regions much larger than that in the gated ones. Therefore, at frequencies close to the frequency of the gated plasmons plasma oscillations in the ungated regions are suppressed, and the effect of the ungated region on the band plasmons can be represented by a frequency-dependent impedance only. The formation of a plasmonic crystal in such structures only requires the electron mean free path in the gated region to exceed its length. There is no such a restriction on the length of the ungated regions as long as decoupling of the gated and ungated plasmons is maintained. We show that driving DC currents induce the tunable peaks in the absorption spectrum of such plasmonic crystals. These peaks are due to the excitation of the band plasmons at the top and the bottom of the plasmonic bandgaps thus demonstrating the fundamental feature of the band energy spectrum. The positions and intensity of these absorption peaks depend on the Mach number and on the ratio of the electron densities in the gated and ungated regions and are controlled by the applied gate and drain bias voltages. This effect could enable excitation of the gap plasmons and facilitate tunable microwave to THz and THz to microwave conversion.

The paper is organized as follows. Section II describes the model approach. Section III presents theoretical analysis and the numerical results for the absorption spectra, and Section IV has discussion of the results and possible applications.

## II. THEORETICAL MODEL

We consider transistor design shown schematically in Fig. 1a. The 2D electron channel of the transistor consists of periodically repeated gated and ungated sections with lengths $L_1$ and $L_2$, respectively. The equilibrium 2D electron density in the gated regions, $n_{01}$, is controlled by the gate voltage while that in the ungated sections, $n_{02}$, remains constant. The system is irradiated by the EM wave of frequency $\omega$ at normal incidence. A DC current bias between the source and the drain contacts of the transistor induces the constant drift velocity $v_0$ in the gated sections.

The EM wave can excite collective electron plasma oscillations in the channel if the frequency $\omega$ matches the plasma eigenfrequencies and $\omega\tau \gg 1$ where $\tau$ is electron momentum relaxation time due to random scattering. The distinct feature of the suggested design is an assumption that $n_{01} \ll n_{02}$ which can be easily achieved using the gate voltage tuning. In this case, the plasma eigenfrequencies in the ungated sections of the channel are well separated from gated plasma modes having much smaller frequencies due to the difference in the equilibrium electron densities and softening of the plasma modes under the gate [30]. This model of the grating-gated electron channel was recently used for theoretical studies of the plasmonic crystal effect on the Dyakonov-Shur instability [23] and formation of the Tamm states in a finite plasmonic crystal [24].

We consider external EM radiation at frequencies close to the frequencies of the gated plasma modes. At these frequencies, the gated sections of the electron channel behave as plasmonic waveguides with ungated sections connecting the neighboring waveguides [23]. A normally incident external EM wave linearly polarized in the direction perpendicular to gate fingers (x-axis) is modulated by the grating gate and produces periodic electric field in the 2D channel: $E(x,t) = E(x)\exp(i\omega t)$ where $E(x) = E(x+L)$ and $L = L_1 + L_2$ is the grating period. In this model, the transistor electron channel can be represented by the equivalent electric circuit diagram shown in Fig. 1b. In this diagram, the gated plasmonic waveguides driven by an external AC electric field are represented by the transmission lines with distributed AC voltage sources $u(x)e^{i\omega t}$ with $u(x) = u(x+L)$. The connecting ungated sections are described by a lumped impedance $\mathcal{R} + i\omega\mathcal{L}$ where $\mathcal{L} = m^*L_2/e^2 n_{02} W$ and $\mathcal{R} = \mathcal{L}/\tau$ are kinetic inductance and resistance, respectively, of the ungated section of width $W$. Here, $m^*$ is an effective electron mass, and $-e$ is electron charge. This $RL$-model is adequate as long as plasma oscillations in the ungated sections remain sufficiently uncoupled from the excited gated plasma modes [23]. In the following, we consider only the homogeneous component of an external AC electric field in the gated sections assuming weak modulation by the grating so that $E(x,t) = E_0 \exp(i\omega t)$. (See more detailed discussion of this approximation in [21].)

Plasma oscillations in the gated sections of the channel can be described by the hydrodynamic equations (equation of continuity and the Euler equation) provided that electron-electron interaction is the dominant source of scattering in the electron system. These equations linearized for the small fluctuations of the electron density $\delta n(x,t)$ and hydrodynamic velocity $\delta v(x,t)$ are

$$\begin{cases} \frac{\partial \delta n}{\partial t} + n_{01} \frac{\partial \delta v}{\partial x} + v_0 \frac{\partial \delta n}{\partial x} = 0 \\ \frac{\partial \delta v}{\partial t} + v_0 \frac{\partial \delta v}{\partial x} + \frac{\delta v}{\tau} = -\frac{e^2}{m^*C} \frac{\partial \delta n}{\partial x} - \frac{e}{m^*} E_0 e^{i\omega t} \end{cases} \quad (1)$$

Here, induced electric potential $\delta V$ is linked to the fluctuation of the electron charge density in the channel as $\delta V = -e\delta n/C$ where $C = \varepsilon \varepsilon_0/d$ is the capacitance per unit area between the channel and the gate, $\varepsilon$ and $d$ are the dielectric constant and the effective thickness of the gate dielectric barrier including the finite thickness of the 2D electron layer, respectively.

Searching solutions of Eqs. (1) in the form $\delta n(\delta v) = \delta n_{q\omega}(\delta v_{q\omega})\exp(-iqx + i\omega t)$ we obtain

$$\begin{cases} (\omega - qv_0)\delta n_{q\omega} - qn_{01}\delta v_{q\omega} = 0 \\ \frac{e^2 q}{m^*C} \delta n_{q\omega} - \left(\omega - qv_0 - \frac{i}{\tau}\right) \delta v_{q\omega} = -\frac{eE_0}{m^*} \end{cases} \quad (2)$$

The system of linear algebraic equations (2) yields expressions for the AC electric current density $\delta j = -e(n_{01}\delta v + v_0 \delta n)$ and induced electric potential $\delta V$ in the gated channel:

$$\begin{cases} \delta j(x,t) = \left(i_1 e^{-iq_1 x} + i_2 e^{-iq_2 x} - \frac{ie^2 n_{01} E_0}{m^*(\omega - i/\tau)}\right) e^{i\omega t} \\ \delta V(x,t) = \frac{1}{\omega C}(i_1 q_1 e^{-iq_1 x} + i_2 q_2 e^{-iq_2 x}) e^{i\omega t} \end{cases} \quad (3)$$

where constant coefficients $i_{1,2}$ are determined by the boundary conditions. Wave vectors $q_{1,2}$ in Eqs. (3) are determined from the quadratic determinantal equation for homogeneous ($E_0 = 0$) system in Eq. (2):

$$(\omega - qv_0)(\omega - qv_0 - i/\tau) - q^2 v_p^2 = 0 \quad (4)$$

Here $v_p = \sqrt{e^2 n_{01}/m^* C}$ is the plasmon velocity in the gated section of the channel. In expression for $\delta j(x)$ in Eq. (3) the first term describes plasmonic contribution and the second one accounts for the Drude contribution to the AC current and in the gated channel.

In the ungated sections, the electric current is conserved: $\delta j(L_1) = \delta j(L_1 + L_2)$, and the electric current and voltage obey the Ohm's law: $V(L_1) - \delta j(L_1)W(\mathcal{R} + i\omega\mathcal{L}) = V(L_1 + L_2)$ where $V(x) = \delta V(x) + u(x)$ is the total electric potential in the gated channel. If only homogeneous component of an external AC electric field in the gated channel is retained, we have $u(0) - u(L_1) = E_0 L_1$. Boundary conditions for solutions (3) follow from these equations and the Bloch conditions in the periodic structure: $\delta j(x+L) = \delta j(x)e^{-ikL}$ and $\delta V(x+L) = \delta V(x)e^{-ikL}$ where $k \in [-\pi/L, \pi/L]$ is the Bloch wave vector. In this model, the boundary conditions take the form

$$\delta j(L_1) = \delta j(0)e^{-ikL} \quad (5)$$

$$\delta V(L_1) - \delta j(L_1)W(\mathcal{R} + i\omega\mathcal{L}) = \delta V(0)e^{-ikL} + E_0 L_1 \quad (6)$$

Substituting Eq. (3) into Eqs. (5) and (6) we obtain the following system of equations for the unknown coefficients $i_{1,2}$:

$$(e^{-iq_1 L_1} - e^{-ikL})i_1 + (e^{-iq_2 L_2} - e^{-ikL})i_2 = \frac{ie^2 n_{01}}{m^*(\omega - i/\tau)} E_0 (1 - e^{-ikL}) \quad (7)$$

$$\left(\frac{q_1}{\omega}(e^{-iq_1 L_1} - e^{-ikL}) - \frac{i\eta\left(\omega - \frac{i}{\tau}\right)L_2}{v_p^2} e^{-iq_1 L_1}\right) i_1$$
$$+ \left(\frac{q_2}{\omega}(e^{-iq_2 L_2} - e^{-ikL}) - \frac{i\eta\left(\omega - \frac{i}{\tau}\right)L_2}{v_p^2} e^{-iq_2 L_2}\right) i_2$$
$$= \left(\frac{L_1}{L_2} + \eta\right) C E_0 L_2 \quad (8)$$

where $\eta = n_{01}/n_{02}$ is the modulation factor. Determinantal equation of this system yields the plasma dispersion equation for the plasmonic crystal formed in the transistor channel. In the "clean" limit ($\omega\tau \to \infty$) we obtain

$$\cos(kL + M\theta) - \cos\theta + \eta(1-M^2)\frac{\omega L_2}{2v_p}\sin\theta = 0 \quad (9)$$

Here $\theta = \frac{\omega L_1}{(1-M^2)v_p}$ and $M = \frac{v_o}{v_p}$ is the Mach number. The last equation agrees with the dispersion equations derived in Refs. [23, 24] if the latter equations are taken in the same limit as Eq. (9).

Fig. 2 presents the numerical solution of Eq. (9). The plots of plasma frequency $f = \omega/2\pi$ as a function of the Bloch vector shown in Fig. 2 demonstrate the plasmonic band spectrum in the first Brillouin zone at several values of the Mach number. Periodic modulation of the electron density results in the gap opening in the plasmonic spectrum. Finite DC bias shifts positions of the gaps in the $k$-space and changes the gap size at any given value of the Bloch vector. In this calculation, we used the values of parameters typical for the InGaAs-based semiconductor structures: $m^* = 0.042 m_e$ ($m_e$ is free electron mass), $\varepsilon = 12.9$, $L_1 = L_2 = 100$ nm, $d = 25$ nm, $n_{02} = 1 \times 10^{16}$ m$^{-2}$, and $\eta = 0.1$. At these values of the material parameters the plasma wave velocity in the gated sections $v_p \approx 3.8 \times 10^5$ m/s so that the Mach numbers $M \leq 0.1$ used in Fig. 2 correspond to the drift velocities $v_0$ well below the saturation values. As seen from Fig. 2, the plasmonic band structure becomes asymmetrical for finite Mach numbers. The asymmetry increases with $M$ with the band gap becoming indirect and dependent on $M$.

All characteristic features of the plasmonic band spectrum can be probed using absorption of an external EM radiation by the band plasmons. Normally incident EM wave interacts with the band plasmons at the center of the first Brillouin zone at $k = 0$. Average electromagnetic power $P$ absorbed in the crystal elementary cell ($0 \leq x \leq L_1 + L_2$) per unit channel width can be found as

$$P = \frac{1}{2}\text{Re}\int_0^{L_1}\delta j(x)\left(-\frac{\partial V^*(x)}{\partial x}\right)dx + \frac{1}{2}|\delta j(0)|^2 \mathcal{R} W \quad (10)$$

where the first term describes the EM power absorbed in the gated section of the elementary cell, and the second one accounts for the EM power absorbed in the ungated section.

In the next section, we solve plasmon hydrodynamic equations and calculate the absorption of the EM radiation by the band plasmons demonstrating tunable resonant behavior.

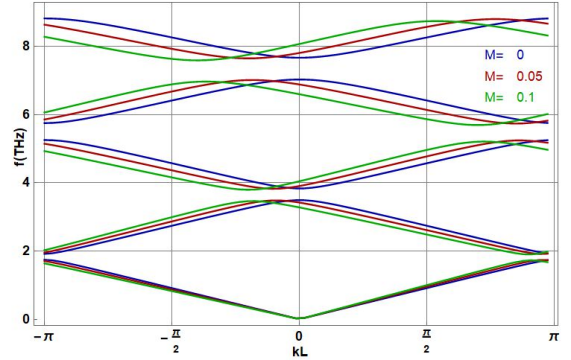

Fig. 2 Plasmonic band spectrum in the first Brillouin zone at Mach numbers $M = 0, 0.05$, and $0.1$ for InGaAs-based periodic semiconductor structures with period $L = 200 nm$, the gate dielectric thickness $d = 25 nm$, and electron densities in the gated and ungated sections $1 \times 10^{15}$ m$^{-2}$ and $1 \times 10^{16}$ m$^{-2}$, respectively. All other parameters are defined in the text.

### III. RESULTS

Normally incident EM wave couples to plasmons at the center of the Brillouin zone, *i.e.* at $k = 0$ in Fig. 2. Plasmonic energy spectrum at the center of the Brillouin zone can be found analytically from Eq. (9) at small Mach numbers $M \ll 1$ and assuming strong modulation $\eta \ll 1$. In this case, the perturbative solution of Eq. (9) in the lowest non-vanishing order in these small parameters yields two series of roots $\omega_{1,2m}$, $m = 1,2,...$:

$$\omega_{1m} = \frac{2\pi m v_p}{L_1} + \frac{\pi m v_p}{L_1}\frac{L_2}{L_1}\left(\sqrt{\eta^2 + \frac{4L_1^2 M^2}{L_2^2}} - \eta\right)$$
$$\equiv \frac{2\pi m v_p}{L_1} + \Delta_{1m} \quad (11)$$

$$\omega_{2m} = \frac{2\pi m v_p}{L_1} - \frac{\pi m v_p}{L_1}\frac{L_2}{L_1}\left(\sqrt{\eta^2 + \frac{4L_1^2 M^2}{L_2^2}} + \eta\right)$$
$$\equiv \frac{2\pi m v_p}{L_1} + \Delta_{2m} \quad (12)$$

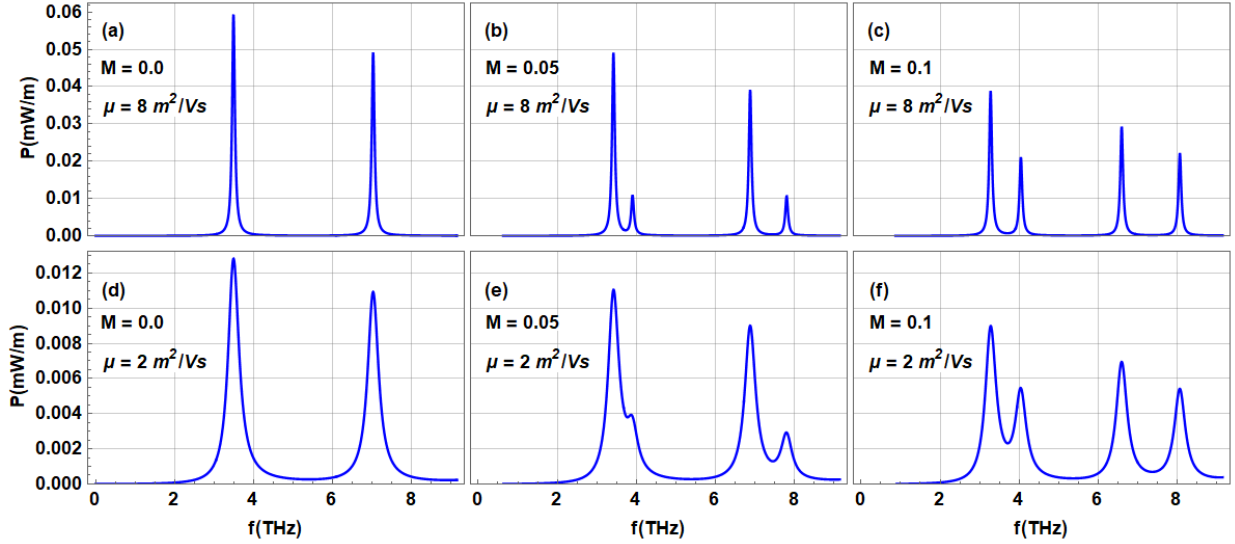

Fig. 3 Electromagnetic power $P$ per unit channel width absorbed in the elementary cell of the InGaAs plasmonic crystal as a function of the frequency $f$ of an external EM radiation at two different values of the electron mobility $\mu$: (a)-(c) $\mu = 8$ m²/Vs; (d)-(f) $\mu = 2$ m² /Vs and different values of the Mach number $M = 0, 0.05, 0.1$. All other parameters are defined in the text.

These roots describe plasmon energies at the top ($\omega_{1m}$) and the bottom ($\omega_{2m}$) boundaries of the energy band gaps opening in the plasmonic crystal spectrum near the frequencies $\omega_m = 2\pi m v_p/L_1$. It is worth noting that the positions of the boundaries as well as the size of the band gap $\Delta_m = \Delta_{1m} - \Delta_{2m}$ depend on the modulation factor $\eta$ and the Mach number $M$ and, therefore, are controllable by the gate voltage and DC bias applied to the transistor.

The closed-form analytical solutions for plasmonic current and voltage distributions as well as the EM power absorption by the plasmons at the center of the Brillouin zone can be readily obtained in the resonant regime at $\omega\tau \gg 1$. In this limit, Eq. (4) yields solutions for the wave vectors $q_{1,2} = (\omega - i/2\tau)/(v_0 \pm v_p)$ corresponding to plasma oscillations propagating in the direction of the DC drift and in the opposite direction, respectively. Such an addition of the plasma and drift velocity was experimentally confirmed in [31]. These solutions are used to find coefficients $i_{1,2}$ in Eqs. (7), (8) and plasmonic contributions to the current and voltage in the gated sections of the channel in Eq. (3). After some cumbersome but straightforward algebra we arrive to the following expressions for the complex plasmonic current $\delta j_m(x)$ and induced voltage $\delta V_m(x)$ in the gated sections ($0 \leq x \leq L_1$) near the $m$-th plasmonic resonance when $\delta\omega = \omega - \omega_m \ll \omega_m$, $M \ll 1$, and $\Omega = \omega - i/2\tau$:

$$\delta j_m(x) = (i_1 e^{-iq_1 x} + i_2 e^{-iq_2 x})$$
$$= \frac{2i\omega e^2 n_{01} v_p E_0}{m^* L_1 \Omega(\Omega - \omega_{1m})(\Omega - \omega_{2m})} \sin\frac{\Omega L_1}{2v_p} \cos\frac{\Omega}{v_p}\left(x - \frac{L_1}{2}\right) \quad (13)$$

$$\delta V(x) = \frac{1}{\omega C}(i_1 q_1 e^{-iq_1 x} + i_2 q_2 e^{-iq_2 x})$$
$$= \frac{2v_p^2 E_0}{L_1 (\Omega - \omega_{1m})(\Omega - \omega_{2m})} \sin\frac{\Omega L_1}{2v_p} \sin\frac{\Omega}{v_p}\left(x - \frac{L_1}{2}\right) \quad (14)$$

Using Eqs. (13) and (14) in Eq. (10) we obtain plasmonic contribution to the absorbed EM power $P_{pl}(\omega)$ at values of $\omega$ close to $\omega_m$:

$$P_{pl}(\omega) = \frac{e^2 n_{01}}{4m^* L_1 \tau} E_0^2 L_1^2 \frac{\delta\omega^2 + 1/4\tau^2}{[(\delta\omega - \Delta_{1m})^2 + 1/4\tau^2][(\delta\omega - \Delta_{2m})^2 + 1/4\tau^2]} \quad (15)$$

Eq. (15) describes the absorption of external EM radiation by the band plasmons near the plasmonic resonances in the crystal elementary cell. This absorption mostly occurs in the gated section of the elementary cell as the contribution of the second term in Eq. (10) into absorption is $\eta$ times smaller than that of the first term.

It follows from Eq. (15) that in the absence of a DC drift ($M = 0$) when $\Delta_{1m} = 0$ the resonant plasmonic absorption occurs only at the bottom boundary of the plasmonic band gap while the plasmonic mode at the top boundary remains optically inactive. These modes are the so called bright and dark plasma modes [7] with different absorption behavior resulting from the different symmetry of the charge and field distributions in these modes and relevant selection rules [21]. An applied DC bias breaks the symmetry of the dark modes, and two peaks corresponding to the plasmon absorption at both boundaries of the plasmonic band gap should appear in the absorption spectrum.

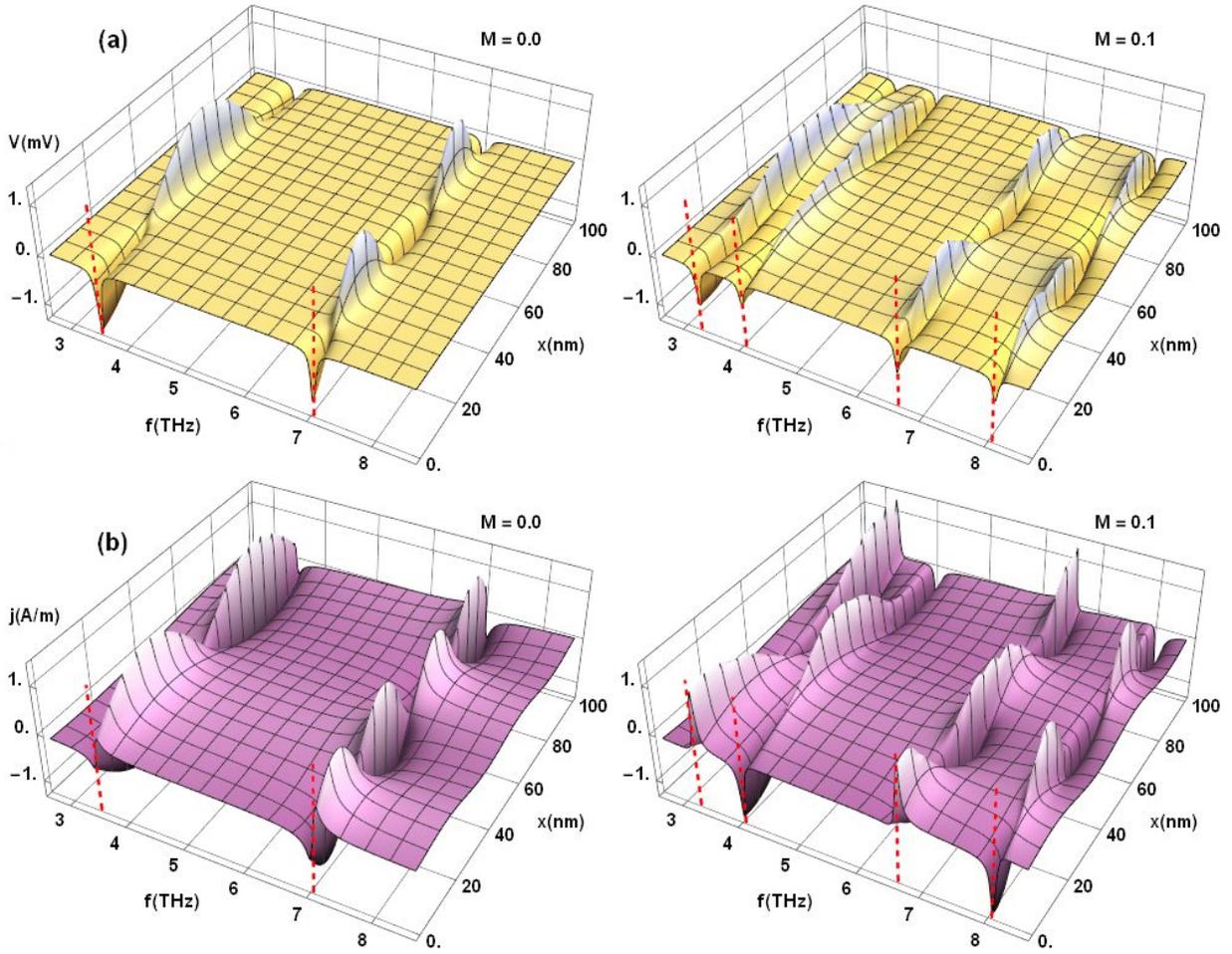

Fig. 4 Spatial distributions of the voltages (a) and the currents (b) in the gated sections of the crystal elementary cell at two different Mach numbers $M = 0, 0.1$ demonstrating excitation of the plasma oscillations at resonant frequencies indicated by dashed lines along $f$-axis. At $M = 0$ plasmons are excited only at the bottom of each energy gap while at finite $M$ plasmons are excited at both boundaries of the energy gaps. In this calculation the mobility $\mu = 8$ m$^2$/Vs was used. All material and geometric parameters are the same as in Figs. 2 and 3.

To describe these features, we numerically solved Eqs. (7) - (9) with boundary conditions in Eqs. (5), (6) and calculated the power absorption spectrum $P(\omega)$ from Eq. (10) free from the restrictions imposed by the perturbation theory used in the analytical solution. In Fig. 3, we plot the EM power per unit channel width absorbed in the elementary cell of the plasmonic crystal as a function of frequency $f$ of the external EM radiation in the frequency interval spanning across first two plasmonic band gaps in Fig. 2. Figs. 3a through 3f show the evolution of the absorption spectrum with increasing Mach number for InGaAs based transistor with mobilities $\mu = 8$ m$^2$/Vs (Figs. 3a-3c) and $\mu = 2$ m$^2$/Vs (Figs. 3d-3f) typical for measurements at cryogenic and room temperatures, respectively. In this calculation we used the value of $E_0 = 1 \times 10^3$ V/m, which is an order of magnitude estimate of the electric field achieved at sub-mW range of impinging THz power for available THz systems [32]. All other parameters are the same as in Fig. 2. As seen from Fig. 3, the additional absorption peak emerges on the high frequency shoulder of the main absorption peak at $M \neq 0$ and increases in amplitude with increasing $M$. This peak is due to plasmonic absorption at the top boundary of the plasmonic band gap as discussed earlier in the text. The amplitudes of the peaks decrease with increasing scattering rates (lower mobilities) but the absorption peaks are still well defined even at room temperatures. In the vicinity of plasma resonances, the absorption is almost entirely determined by the plasmons while the Drude absorption is very small. Drude absorption becomes prevalent when $\omega \to 0$ and is not seen in the frequency scale of Fig. 3.

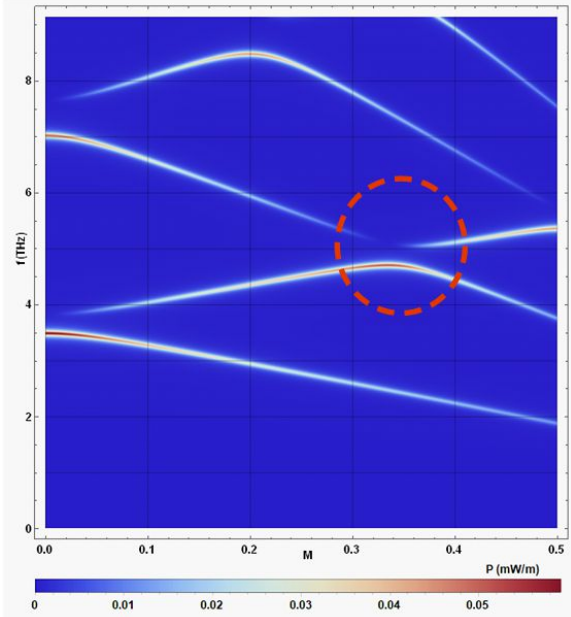

Fig. 5 Power absorption map in the $(\omega, M)$ plane showing evolution of the absorption peaks for the first two energy band gaps with increasing Mach number $M$. Additional anti-crossings accompanied by the appearance of new bright and dark modes occur when the absorption peaks belonging to the different band gaps approach each other, see the encircled anti-crossing corresponding to the AMS effect measured in [18].

In Fig. 4, we plot spatial distributions of the voltages (Fig. 4a) and the currents (Fig. 4b) in the gated section of the crystal elementary cell at different frequencies and DC biases demonstrating excitation of the plasma oscillations at resonant frequencies indicated by the dashed lines on the plots. At $M = 0$, plasmons are excited at the bottom of the plasmonic gap only while at finite $M$ plasma excitations appear at both boundaries of the plasmonic gap. Spatial distribution of the currents and voltages roughly follows Eqs. (13) and (14).

In Fig. 5, the dissipated power $P$ is mapped in the $(\omega, M)$ plane illustrating a high tunability of the plasmonic absorption by a DC bias. This figure shows evolution of the absorption peaks for the first two energy band gaps when $M$ changes in the interval $0 \leq M \leq 0.5$. The positions of the peaks as well as the size of the energy gaps strongly depend on $M$. At larger $M$ ($M \approx 0.3$) the absorption peaks belonging to different band gaps approach each other resulting in the mode anti-crossings indicated in Fig. 5 by a dashed line circle. These anti-crossings are accompanied by the appearance of the bright and the dark modes as seen in Fig. 5 and present another manifestation of the AMS effect when the plasma resonant peak experiencing redshift with increasing current changes to the blue shifting peak [18, 19].

The presented results demonstrate that the absorption spectrum of the current-biased plasmonic crystal is rich in different features tunable by the applied DC bias. These features are more pronounced at cryogenic temperatures but persist up to the room temperatures opening an opportunity for designing frequency sensitive tunable detectors of the THz EM radiation.

## IV. DISCUSSION AND CONCLUDING REMARKS

We investigated a plasmonic crystal structure designed to optimize the quality factor of plasmonic resonances by connecting the resonant gated regions via ungated regions with large electron densities and elevated plasmonic frequencies decoupled from the resonances in the gated regions. This design aims to achieve resonant behavior at room temperature across various material systems, including graphene, III-V and III-N semiconductor materials, and p-diamond due to the short length of the individual gated plasmonic cavities compared to the electron mean free path. The flexibility in the length of the ungated region allows the unit cell length to be chosen sufficiently long for an optimum coupling between the impinging THz radiation and the plasmonic crystal with the metal grating acting as a distributive resonant antenna tunable by choosing the appropriate length of the ungated regions.

A significant feature of the analyzed plasmonic crystal spectrum is the mode anti-crossings with tunable gaps in the plasmonic spectrum that could be modulated by the gate or the current biases. The spectrum of the plasmonic crystals with anti-crossings is similar to that of the bilayer graphene [33] or the narrow gap semiconductors [34]. Near the band boundaries the spectrum becomes parabolic and could be characterized by a plasmonic effective mass like that of a roton [35] and dependent on the gate and current biases. At finite Mach numbers, the tunable gaps become indirect with this feature being more pronounced at high Mach numbers.

At finite Mach number plasma states become optically active at both boundaries of the plasmonic band gap. This opens an opportunity for excitation of the plasmons forming the gap using power pumping at the gap plasmon frequency, *i.e.* transfer of power from a lower to a higher frequency signal. This method can potentially be used for the microwave to THz conversion.

At sufficiently large Mach numbers the plasmonic crystal could become unstable [16, 17, 23]. The plasmonic crystal design considered in this paper makes it easier to achieve the resonant conditions required for such an instability because it does not require the resonant conditions for the ungated sections. In contrast to similar instabilities in a single plasmonic FET, such an instability in a plasmonic crystal should be orders of magnitude more powerful. This advantage of the analyzed design also applies to the plasmonic boom instability that should occur for $M>1$ [14].

Other effects to be investigated include the dependences of the plasmonic crystal response on the impinging radiation angle and polarization, and galvanomagnetic effects. All these effects should be much easier to realize for the proposed design allowing for the optimization of the unit cell length.

Implementing a plasmonic crystal increases the active area of a THz or sub-THz device by orders of magnitude compared to a single plasmonic TeraFET. A large active area of the plasmonic crystal in comparison to a single plasmonic TeraFET makes it uniquely suited for gas, fluid, proximity, and biomedical sensors. The commensurate increase in sensitivity for detector applications and generated power for plasmonic THz oscillations makes plasmonic crystals to be prime candidates for 6G and beyond THz communication systems, the line-of-sight THz detectors, biomedical and industrial sensing and imaging, THz frequency conversion systems, and advanced plasmonic devices for radar, security, and defense applications.